\RequirePackage[T1]{fontenc}
\documentclass[12pt]{article}

\usepackage[height=8.85in,width=6.45in]{geometry}

\usepackage[utf8]{inputenc}
\usepackage{amsmath}
\usepackage{amssymb}
\usepackage{mathtools}
\numberwithin{equation}{section}
\usepackage{slashed}
\usepackage{braket}
\usepackage{hyperref}
\usepackage{cite}
\usepackage{times}
\usepackage{courier}
\usepackage{bm}
\usepackage{subfig}

\usepackage{xcolor}
\usepackage{mdframed}

\renewenvironment{figure}[1][]{
  \begin{originalfigure}[#1]
    \begin{mdframed}[linecolor=black!0,backgroundcolor=black!1]
}{
    \end{mdframed}
  \end{originalfigure}
}
\def\Nequals#1{$\mathcal{N}{=}#1$}
\def\bA{\mathbb{A}}
\def\cA{\mathcal{A}}
\def\cF{\mathcal{F}}
\def\cM{\mathcal{M}}
\def\cO{\mathcal{O}}
\def\cU{\mathcal{U}}
\def\SO{\mathrm{SO}}
\def\SU{\mathrm{SU}}
\def\U{\mathrm{U}}
\def\tr{\mathop{\mathrm{tr}}}
\let\met\mathsf
\def\vev#1{\langle#1\rangle}

\usepackage{slashed}

\def\CL{{\cal L}}

\def\CN{{\cal N}}

\def\U{\mathrm{U}}
\def\SU{\mathrm{SU}}

\def\SO{\mathrm{SO}}

\def\Spin{\mathrm{Spin}}

\def\SR{\mathsf{R}}
\def\SC{\mathsf{C}}
\def\SA{\mathsf{A}}
\def\SF{\mathsf{F}}
\def\SD{\mathsf{D}}

\def\beq#1\eeq{\begin{align}#1\end{align}}

\begin{document}
\begin{titlepage}

\begin{flushright}
IPMU-17-0140
\end{flushright}

\vskip 3cm

\begin{center}

{\Large \bfseries Anomalies involving the space of couplings \\[.5em]
and the Zamolodchikov metric}

\vskip 1cm
Yuji Tachikawa and Kazuya Yonekura
\vskip 1cm

\begin{tabular}{ll}
 & Kavli Institute for the Physics and Mathematics of the Universe, \\
& University of Tokyo,  Kashiwa, Chiba 277-8583, Japan
\end{tabular}

\vskip 1cm

\end{center}

\noindent
The anomaly polynomial of a theory can involve not only curvature two-forms of the flavor symmetry background but also two-forms on the space of coupling constants.
As an example, we point out that there is a mixed anomaly between the R-symmetry and the topology of the space of exactly marginal couplings of class S theories.
Using supersymmetry, we translate this anomaly to the K\"ahler class of the Zamolodchikov metric.
We compare the result against a holographic computation in the large $N$ limit.

In an appendix, we explain how to obtain the bosonic components in the supergravity completion of the curvature squared terms using compatibility with the topological twist.

\end{titlepage}

\tableofcontents

\section{Introduction and summary}
The anomaly polynomial $\bA_{D+2}$ of a $D$-dimensional quantum field theory encodes the anomalous variation of the phase of the partition function in the presence of the background fields in terms of the well-known descent procedure.
We often think of $\bA_{D+2}$ as a functional of the background gauge fields and of the background metric,
but it is known from the early days of the study of anomalies that it can also depend on the topology of the space of couplings \cite{Moore:1984dc,Moore:1984ws,Manohar:1984zj}.\footnote{Strictly speaking in these papers the scalars were considered dynamical. But before 't Hooft \cite{tHooft:1979rat} the gauge fields in the anomalies were also mostly considered dynamical.
In this sense the anomaly involving the space of couplings is  known from the early days.}

One example can be constructed as follows. Take an arbitrary target space $\cM$ and a $\U(N)$ gauge field $\cA$ on it. 
Consider a $D$-dimensional free theory of a dynamical chiral fermion $\psi$ in the fundamental of $\U(N)$ coupled to  a background scalar $\phi$ which is a map to $\cM$ in the following manner:
the fermion $\psi$ is minimally coupled to the pull-back under $\phi$ of the $\U(N)$ field $\cA$.
Of course the theory has the anomaly polynomial \begin{equation}
\bA_{D+2}=\hat A(TX) \tr e^{i\phi^*(\cF) / 2\pi}
\end{equation} where $X$ is the worldvolume of the theory and $\cF$ is the curvature two-form on $\cM$.
When the scalar $\phi$ is considered dynamical, 
this is known under the name of the $\sigma$-model anomaly,
and makes the theory ill-defined when not canceled. 
When $\phi$ is considered as a background field, 
this anomaly polynomial involves differential forms on the space $\cM$ of couplings,
and serves as one of the characteristic properties of the theory,
just as the ordinary 't Hooft anomalies do.

To readers who found the example above rather artificial, let us provide a more meaningful case.
In \cite{Gaiotto:2009we} Gaiotto introduced a large class of 4d \Nequals2 theories obtained by compactifying a 6d \Nequals{(2,0)} theory on a Riemann surface $C$, possibly decorated with punctures. 
Here let us consider a simple case where $C$ is of genus $g$ without any puncture.
This construction gives rise to a family of 4d \Nequals2 superconformal field theories (SCFTs), now known as class S theories,
whose space of exactly marginal couplings is the moduli space $\cM_g$ of the genus-$g$ Riemann surfaces.
We show below that this theory has a hitherto-unappreciated term in the anomaly polynomial of the form \begin{equation}
\bA_6 \supset  [\frac{\omega}{2\pi}] P \label{schematic}.
\end{equation} 
Here, $[\omega]$ is a degree-2 cohomology class on $\cM_g$
and $P$ is a certain degree-four cohomology class constructed from 
the background R-symmetry gauge field and the spacetime metric.
The anomaly \eqref{schematic} can be determined from the known anomaly polynomial of the 6d \Nequals{(2,0)} theory \cite{Harvey:1998bx,Intriligator:2000eq,Yi:2001bz,Ohmori:2014kda}.

Now, a mixed anomaly between the Weyl transformation and the K\"ahler transformation of the space of exactly marginal couplings of general 4d \Nequals2 SCFTs was described in \cite{Gerchkovitz:2014gta,Gomis:2014woa,Gomis:2015yaa}.\footnote{%
There they conjectured that the K\"ahler potential would be globally well defined but this was answered negatively in a recent paper \cite{Donagi:2017vwh}. 
This latter paper  actually gave the impetus of the investigation which led to this short note.}
When one reads their derivation carefully, one finds that their analysis already implies an anomaly of the form \eqref{schematic} above,
with an added bonus that $[\omega]$  is proportional to the cohomology class $[\omega^\text{Z}]$ of the K\"ahler form $\omega^\text{Z}$ of the Zamolodchikov metric.
Turning the logic around, this means that we can easily fix $[\omega^\text{Z}]$ in terms of the anomaly polynomial of the 6d \Nequals{(2,0)} theory.

Finally, we note that the Zamolodchikov metric and therefore $\omega^\text{Z}$ is computable in the large $N$ limit by means of the AdS/CFT correspondence \cite{Maldacena:1997re} using the holographic dual of the class S theories \cite{Maldacena:2000mw,Gaiotto:2009gz}.
This is known to be proportional to the standard Weil-Petersson metric on $\cM_g$,\footnote{%
The author does not know who originally noticed this; he forgot from whom he first learned the fact. 
This is surely a common knowledge among those who study class S theories using AdS/CFT.}
but the precise proportionality coefficient has not been computed to the author's knowledge.
We will show below that the class of the K\"ahler form computed holographically is 
compatible with the computation from the anomaly as above,
using a classic mathematical result by Wolpert \cite{Wolpert1,Wolpert2}.

The rest of the note is devoted to implement the computations outlined above:
in Section~\ref{sec:2} we compute  $\bA_6$ of the class S theory from the anomaly of the 6d \Nequals{(2,0)} theory and then use it to compute the K\"ahler class $[\omega^\text{Z}]$ of the Zamolodchikov metric. 
Then in Section~\ref{sec:3} we compute the Zamolodchikov metric using holography,
determine the proportionality coefficient with respect to the standard Weil-Petersson metric,
and compare it against the result in Section~\ref{sec:2}.

In \hyperref[sec:app]{the appendix}, we explain how one can obtain the bosonic components in the supergravity completion of the curvature square terms by imposing the compatibility with the topological twist. 
This gives another way to confirm the required coefficients used in this paper,
independent from the standard supergravity computations.

\section{Field theoretical computations}
\label{sec:2}
\subsection{The 4d anomaly from the 6d anomaly}
We start from the anomaly polynomial \cite{Harvey:1998bx,Intriligator:2000eq,Yi:2001bz,Ohmori:2014kda} of the 6d \Nequals{(2,0)} theory of type $G =A_{n-1}, D_n, E_{6,7,8}$: \begin{equation}
\bA_8 =\frac{h^\vee_G  d_G }{24}p_2(NY) +\frac{r_G }{48}\left(p_2(NY)-p_2(TY)+\frac14(p_1(NY)-p_1(TY))^2 \right).
\end{equation}
Here $Y$ is the worldvolume of the theory, $TY$ is its tangent bundle,
$NY$ is the $\SO(5)$ R-symmetry bundle,
and $p_1$, $p_2$ are the Pontryagin classes;
$h^\vee_G $, $d_G $ and $r_G $ are the dual Coxeter number, the dimension and the rank of the Lie algebra of type $G $.
We use the convention that the anomaly polynomial $\bA_{D+2}$ gives the $(D+1)$-dimensional Chern-Simons invariant $\exp(\int_{Y_{D+1}} 2\pi i \bA_{D+1})$, where 
$\bA_{D+2}=d\bA_{D+1}$.

The \Nequals2 class S theory of our interest is obtained by compactifying the 6d theory on a Riemann surface $C$ of genus $g$ without any punctures
so that we introduce a nonzero curvature to the subgroup $\SO(2)\subset \SO(5)$ of the R-symmery
which cancels the curvature of $C$.
This means that the Chern roots of $NY$ is $\pm 2\alpha$ and $\pm t$
where $\pm \alpha$ are the Chern roots of the $\SU(2)_R$ background field
and $t$ is the first Chern class of the tangent bundle of $C$.

Using $p_1=\sum_i \lambda_i^2$ and $p_2=\sum_{i<j}\lambda_i^2 \lambda_j^2$ 
when the Chern roots are $\pm\lambda_i$,
we easily get \begin{equation}
\bA_6 \supset -\left[\frac{h^\vee_G  d_G }{6}c_2+\frac{r_G }{12}( c_2 +\frac{p_1}{4})\right] \int_C t^2.
\label{classS}
\end{equation}
Here, $p_1$ is the Pontryagin class of the spacetime metric
and $c_2$ is the second Chern class of the $\SU(2)_R$  symmetry.

We note that $\int_C t^2$ is understood as follows: 
$t$ is considered as the first Chern class of the relative tangent bundle of $C$ over $\cM_g$
(i.e.~the tangent bundle of the universal bundle $\cU$ where $C\hookrightarrow \cU\twoheadrightarrow \cM_g$ minus the pull back of the tangent bundle of $\cM_g$).
Then $t^2$ is a 4-form on $\cU$, and we obtain a 2-form on $\cM_g$ by integrating over the fiber $C$.

\subsection{Finding the K\"ahler potential}
Suppose now that a given 4d \Nequals2 SCFT has a space of exactly marginal couplings
parameterized by $\cM$ with local complex coordinates $\tau^I$.
We  normalize the corresponding exactly marginal operators so that they enter in the deformation of the Lagrangian as
\begin{equation}
\frac1{\pi^2}\int d^4 x (\tau^I \cO_I+\bar\tau^{\bar J} \bar \cO_{\bar J})\label{poo}
\end{equation}
following the convention of \cite{Gerchkovitz:2014gta,Gomis:2014woa,Gomis:2015yaa}.
We then define the Zamolodchikov metric $g_{I\bar J}^\text{Z}$ by the formula
\begin{equation}
\vev{\cO_I(x) \bar \cO_{\bar J}(0)}=\frac{g_{I\bar J}^\text{Z}}{x^8}.
\end{equation} 
This is known to be K\"ahler: $g_{I\bar J}^\text{Z}=\partial_I \bar\partial_{\bar J} K$.

In \cite{Gomis:2015yaa} the authors identified that the K\"ahler transformation $K\mapsto K+F+\bar F$ needs to be accompanied by a shift of the counter terms \begin{equation}
S\mapsto S+\frac{1}{192\pi^2}\int d^4x d^4\theta \mathcal{E} \frac{F}2 (\Xi-W^{\alpha\beta}W_{\alpha\beta})+c.c. \label{foo}
\end{equation}
where $\Xi-W^{\alpha\beta}W_{\alpha\beta}$ is the supersymmetric completion of the Euler density constructed in \cite{Butter:2013lta}.
Using the component expansion given in (5.5) of \cite{Butter:2013lta}, we see
that this shift contains the terms of the form
\begin{multline}
S\mapsto S-
\frac{1}{192\pi^2}\int d^4x 
\frac{F+\bar F}{2}(
\frac 12 \SC^{abcd} \SC_{abcd}-\SR_{ab}\SR^{ab}+\frac13 \SR^2
) \\
-\frac{1}{192\pi^2}\int d^4x 
\frac{F-\bar F}{2}
\left(
\frac12 \SR^{abcd}\tilde \SR_{abcd}
+ \frac12 R(V)_{ab}{}^i_j \tilde R(V)^{ab}{}^j_i \right)
\end{multline} 
where 
$\SC_{abcd}$, $\SR_{ab}$ and $\SR$ are the Weyl tensor, Ricci tensor and Ricci scalar, respectively, and
$R(V)_{ab}{}^i_j$ is the background gauge field strength of the $\SU(2)_R$ symmetry in their convention, see (B.2) and (B.3) of \cite{Butter:2013lta}.
We find that $K\mapsto K+F+\bar F$ is accompanied by\begin{equation}
S\mapsto S-
 \int \frac{F+\bar F}2 \frac{e}{12}
 -\int \frac{F-\bar F}2(\frac{c_2}{6}+ \frac{p_1}{24}) \label{bar}
\end{equation}
where $e$ is the Euler density.\footnote{%
In \hyperref[sec:app]{the appendix}, 
we give an independent derivation of the combinations of the bosonic terms in $\int d^4\theta\mathcal{E}(\Xi-W^{\alpha\beta}W_{\alpha\beta})$ in \eqref{bar} 
and $\int d^4\theta\mathcal{E} W^{\alpha\beta}W_{\alpha\beta}$ in \eqref{baz}.
}

The K\"ahler form of the Zamolodchikov metric is \begin{equation}
\omega^\text{Z}=i \partial \bar\partial K = -idA
\qquad\text{where}\qquad
A=\frac12(\partial-\bar\partial)K.
\end{equation} Therefore, $K\mapsto K+F+\bar F$ does \begin{equation}
A\mapsto A+\frac12(\partial F-\bar\partial\bar F)=A+d(\frac{F-\bar F}{2}).
\end{equation} This means that the shift of the counter term \eqref{bar} is exactly the gauge variation one obtains from the anomaly polynomial \begin{equation}
\bA_6 \supset  -[\frac{\omega^\text{Z}}{2\pi}](\frac{c_2}{6}+ \frac{p_1}{24}).\label{one}
\end{equation} 

This is however not the whole story. 
In the \Nequals2 supergravity, there is another chiral term one can write: \begin{multline}
\frac{1}{192\pi^2}\int d^4x d^4\theta \mathcal{E} \frac{G}2 W^{\alpha\beta}W_{\alpha\beta} +c.c. \\
=\frac{1}{192\pi^2}\int d^4x \frac{G+\bar G}{2}(\frac12 \SC^{abcd} \SC_{abcd} +\frac1{4}R(V)_{ab}{}^i_j R(V)^{ab}{}^j_i ) + \int \frac{G-\bar G}2(\frac{c_2}{12}+\frac{p_1}{24}),
\label{baz}
\end{multline} which can produce another term in the anomaly polynomial of the form
\begin{equation}
\bA_6 \supset -[\frac{\omega^X}{2\pi}] ( \frac{c_2}{12}+\frac{p_1}{24})\label{two}
\end{equation}
where $\omega^X$ is a certain closed 2-form on the space of couplings.

Let us specialize to the case of the class S theory on a genus-$g$ Riemann surface.
Matching \eqref{classS} against the sum of \eqref{one} and \eqref{two}, one finds 
\begin{equation}
[\frac{\omega^\text{Z}}{2\pi}]=(2h^\vee_G d_G+\frac{r_G}2)\int_C t^2,\qquad
[\frac{\omega^\text{X}}{2\pi}]=-2h^\vee_G d_G \int_C t^2.
\end{equation}
Choosing the type to be $A_{N-1}$ and taking the large $N$ limit, we find 
 \begin{equation}
[\frac{\omega^\text{Z}}{2\pi}]\sim 2N^3\int_C t^2 \label{yoo}
\end{equation} as cohomology classes.

\section{Holographic computations}
\label{sec:3}
\paragraph{Weil-Petersson metric:}
First let us recall the Weil-Petersson metric on the moduli space $\cM_g$ of the genus-$g$ Riemann surface.
Given Beltrami differentials $\mu_I$ on a Riemann surface $C$, 
we define the Hermitean structure on them by \begin{equation}
g_{I\bar J}^\text{WP}=\int_C \mu_I \bar \mu_{\bar J} dA
\end{equation} where $dA$ is the area form of curvature $-1$, so that $\int_C dA=4\pi(g-1)$.
Then the K\"ahler form of the Weil-Petersson metric is given by \begin{equation}
\omega^\text{WP}=g_{I\bar J}^\text{WP} ( i d\tau^I \wedge d\bar \tau^{\bar J} ). 
\end{equation}
It is a classic mathematical result by Wolpert \cite{Wolpert1,Wolpert2} that the relation\begin{equation}
\frac{\omega^\text{WP}}{2\pi^2}=\int_C t^2
\end{equation} 
holds as differential forms, not just as cohomology classes.

\paragraph{Maldacena-Nu\~nez solution:}
The holographic dual of the class S theory of type $A_{N-1}$ on a genus-$g$ Riemann surface
is given in \cite{Maldacena:2000mw,Gaiotto:2009gz}.
The 11d metric is of the form
\begin{multline}
ds^2_{11}=(\pi N \ell_p^3)^{2/3}W^{1/3} \Bigl[ 2 ds^2_{AdS_5} +  ds^2_{H_2/\Gamma} + \\
d\theta^2 + W^{-1} \{ \cos^2\theta(d\psi^2+ \sin^2\psi d\phi^2) + 2\sin^2\theta(d\chi+A)^2 \} \Bigr].
\end{multline}
where $ds^2_{AdS_5}$ is the $AdS_5$ metric of unit radius,
$ds^2_{H_2/\Gamma}$ is the metric on the Riemann surface $C$ represented as a quotient of the Poincar\'e disk of curvature radius 1 by a discrete group $\Gamma$,
the coordinates $\theta$, $\psi$, $\phi$ and $\chi$ parameterize the internal space which is topologically of the form $S^4$, and
$W:=1+\cos^2\theta$. 
Our convention is that the 11d Euclidean action is \begin{equation}
S_{11} \supset  -  \frac{1}{16\pi G_N^{(11)}} \int d^{11} x \sqrt{\met g} \met R,\qquad G_N^{(11)}=16\pi^7 \ell_p^9.
\end{equation}

The point is that the deformation of $C=H/\Gamma$ is parameterized by $\cM_g$,
the moduli space of genus-$g$ Riemann surfaces.
The space $\cM_g$ appears as the target space of the massless scalars in five dimensional supergravity,
which in turn can be identified with the space of exactly marginal couplings of the dual SCFT.

\paragraph{Reduction to five dimensions:}
Let us proceed with our computation.
We normalize the 7d metric as
%\begin{equation}
%ds^2_7=(\pi N\ell_p^3)^{2/3} (2ds^2_{AdS_5} + ds^2_{H_2}).
%\end{equation} 
\begin{equation}\label{eq:normalization}
ds^2_7= (2ds^2_{AdS_5} + ds^2_{H_2/\Gamma}).
\end{equation} 
Integrating over the four-sphere part, we get the 7d action 
%\begin{equation}
%S_7 \supset  \frac{1}{16\pi G_N^{(7)}} \int d^{7} x \sqrt{\met g} \met R +\cdots
%\end{equation} 
\begin{equation}
S_7 \supset  - \frac{1}{16\pi {G}_N^{(7)}} \int d^{7} x \sqrt{\met g} \met R +\cdots
\end{equation} 
where 
%\begin{equation}
%\frac{1}{G_N^{(7)}} = \frac{1}{G_N^{(11)}} (\pi N \ell_p^3)^2 \frac{8}{3}\pi^2 \sqrt{2}.
%\end{equation}
\begin{equation}
\frac{1}{G_N^{(7)}} = \frac{1}{G_N^{(11)}} (\pi N \ell_p^3)^3 \frac{8}{3}\pi^2 \sqrt{2} = \frac{N^3}{3\sqrt{2} \pi^2}.
\end{equation}

We now reduce it further to five dimensions, including the deformation of the Riemann surface.
In general, under a small deformation $\met g\mapsto \met g + \met h$, we have \begin{equation}
\int \sqrt{ \met g} \met R \mapsto \int \sqrt{ \met g} \left(   \met R - \frac{1 }{4}    \nabla_\mu \met h_{\nu\rho} \nabla^\mu \met h^{\nu\rho} + \cdots \right),
\end{equation}
where we neglected terms which do not contribute to the following computation.
Now note that the Beltrami differential $\mu:=\tau^I \mu^z_{\bar z}{}_I$ deforms the internal metric as \begin{equation}
|dz|^2 \mapsto |dz+\mu  d\bar z|^2
\end{equation} which means that \begin{equation}
\met g \mapsto \met g + \met h, \qquad  \met g = |dz|^2, \qquad  \met h:= \mu (d\bar z)^2+\bar \mu (dz)^2  ,  %2 (\mu + \bar \mu)
\end{equation} %where the indices are appropriately placed. 
up to an overall conformal factor. 
This means that upon reduction to 5d one finds \begin{equation}
S_5 \supset - \frac{4\pi (g-1)}{16\pi G_N^{(7)}} \int d^5 x \sqrt{\met g} \met R +
\frac{ 2g_{I\bar J}^\text{WP}}{16\pi G_N^{(7)}}
 \int d^5 x\sqrt{\met g} \met g^{\mu\nu} \partial_\mu \tau^I \partial_\nu \bar\tau^{\bar J}
\end{equation}

\paragraph{Translation to the dual SCFT:}
At this point, we can use the formula \cite{Henningson:1998gx} for the central charge $a\sim c$ to compute
\begin{equation}
c=\frac{\pi R_{AdS_5}^3}{8G_N^{(5)}} = \frac{N^3}{3}(g-1)
\end{equation}    where $1/G_N^{(5)} = 4\pi (g-1)/G_N^{(7)}$ and $R_{AdS_5}=\sqrt{2}$ in the above normalization of the metric \eqref{eq:normalization}.
This reproduces the standard result.

We are more interested in the Zamolodchikov metric on $\cM_g$,
for which we use the formula for the two-point function under AdS/CFT given in \cite{Freedman:1998tz}.
The formula says that given the action \begin{equation}
S_5 \supset  \frac{\eta}2 \int d^5 x\sqrt{\met g} \met g^{\mu\nu} \partial_\mu \phi \partial_\nu \phi,
\end{equation} for a real scalar $\phi$ in five dimensions,
the corresponding operator has the two-point function \begin{equation}
\vev{\underline{\cO}(x)\underline{\cO}(0)}=\eta R_{AdS_5}^3 \frac{24}{\pi^2} \frac1{x^8},
\end{equation} with the caveat that the deformation is introduced via the coupling \begin{equation}
S\mapsto S+\int d^4x \phi \underline{\cO}
\end{equation} \emph{without} an additional factor of $\pi^2$ in the denominator as in \eqref{poo}.

Carefully collecting all the factors, one finds \begin{equation}
g_{I\bar J}^\text{Z}=\frac{2g_{I\bar J}^\text{WP}}{16\pi G_N^{(7)}} R_{AdS_5}^3 \frac{24}{\pi^2}  (\pi^2)^2
=2N^3\frac{g_{I\bar J}^\text{WP}}{\pi} 
\end{equation}meaning that \begin{equation}
\frac{\omega^\text{Z}}{2\pi} \sim 2N^3 \frac{\omega^\text{WP}}{2\pi^2} = 2N^3 \int_C t^2
\end{equation} in the large $N$ limit.
This is consistent with the result \eqref{yoo} we obtained from the consideration of the  anomaly in the last section.

\section*{Acknowledgments}
YT thanks D. R. Morrison for discussions,
and T. Maxfield for pointing out a minor error in (2.2) in  an older version of the manuscript.
YT is partially supported in part by JSPS KAKENHI Grant-in-Aid (Wakate-A), No.17H04837 
and JSPS KAKENHI Grant-in-Aid (Kiban-S), No.16H06335, and 
also supported in part by WPI Initiative, MEXT, Japan at IPMU, the University of Tokyo.
KY is supported in part by the WPI Research Center Initiative (MEXT, Japan), and also supported by JSPS KAKENHI Grant-in-Aid (17K14265).

\appendix

\section{Supersymmetric completion of $R^2$ terms via holomorphy}
\label{sec:app}

In \cite{Witten:1995gf}, Witten studied the structure of local actions in topologically twisted $\CN=2$ supersymmetric theory.
In particular, the terms of the form
\beq
F(u) {\rm [gravity]}
\eeq
were studied, where $F(u)$ represents dimensionless functions of the bottom components of vector multiplets $u$,
and ${\rm [gravity]}$ represents dimension-four operators constructed from background gravity, such as the Euler density and signature density.

Here we use the result of \cite{Witten:1995gf} to determine the allowed forms of the supersymmetric terms for more general background fields in the physical theory. 
We consider the background metric, the background field $\SA$ for $\SU(2)_R$ symmetry, and
the background auxiliary field $\SD$ which is coupled to the bottom component of the energy-momentum multiplet.\footnote{
The R-symmetries contain a subgroup $\U(1) \subset \U(1)_R \times \SU(2)_R$ which is non-R in terms of the $\CN=1$ subalgebra. 
This gives an $\CN=1$ current multiplet, and hence there exists an $\CN=1$ background vector multiplet coupled to it.
The $\SD$ is the $D$-component of this vector multiplet.}
The topological twist means that the gauge field $\SA$ is taken to coincide with the spin connection for the right-handed
spinor of the Lorentz group $\Spin(4) = \SU(2)_\ell \times \SU(2)_r$. Also, the auxiliary field $\SD$ is taken to be
proportional to the Ricci scalar $\SR$ as $\SD=\alpha \SR$ for some numerical constant $\alpha$ which is independent of the theory.\footnote{
One way to see this is as follows. In conformal supergravity formulation, the Ricci scalar $\SR$ gives a coupling 
to scalar fields $\phi$ which is proportional to $\SR |\phi|^2$ (or more generally $|\phi|^2$ is replaced by the Kahler potential).
However, we want to have massless Coulomb branch fields on any background. Therefore, we have to introduce $\SD$
to cancel the mass term induced by $\SR$. }

This computation reproduces the combinations found in \cite{Butter:2013lta}, so the final results in this appendix are not new.
But it is fun to see how the holomorphy of the topologically twisted theory arises from the physical theory, and how it reproduces the supergravity computations.

\paragraph{Derivation of the allowed action:}
 It was argued in \cite{Witten:1995gf}
that after the topological twist, the action of the above form for constant $u$ is holomorphic with respect to $u$.
Also, the original physical theory must satisfy reflection positivity in Euclidean space which means that parity-odd terms are imaginary and parity-even terms are real in the Euclidean action. 
From these conditions, we impose the ansatz for the Euclidean action as
\beq
- \CL  &\ni ( f + \bar{f})e + ( g - \bar{g}) p_1 + ( h - \bar{h}) c_2 \nonumber \\
& + ( j +\bar{j}) [\SR_{\mu\nu\rho\sigma}^2] + (k +\bar{k}) [\SR^2] +  ( l +\bar{l}) [\SF^2] + (m +\bar{m})[\SD^2].
\eeq
Here, the $f ,g, h, j, k , l,m$ are holomorphic functions of $u$, $\SF=d \SA +\SA^2$ is the field strength of the $\SU(2)_R$ gauge field $\SA$, and $\SR_{\mu\nu\rho\sigma}$
is the Riemann curvature, and $\SD$ is the auxiliary field mentioned above. We remark that $R(V)$ in the main text is related to $\SF$ as $R(V)=2\SF$.
In the action, we neglect a term proportional to $\nabla^2 \SR$ because it is a total derivative when $u$ is constant and hence cannot be determined by the method of this appendix.

The $e, p_1, c_2, [\SR_{\mu\nu\rho\sigma}^2] ,[\SR^2], [\SF^2], [\SD^2]$ are defined as follows.
Let 
\beq
\hat{\SR} &=\frac{1}{2 \pi} \left( \frac{1}{2}\SR_{ij \mu \nu} dx^\mu \wedge dx^\nu  \right)_{ij}, & 
\hat{\SF} &= \frac{i}{2\pi} \left( \frac{1}{2}\SF_{a \mu \nu} \left(  -i\frac{\sigma^a}{2} \right)  dx^\mu \wedge dx^\nu  \right)
\eeq 
be the Riemann curvature $\SR$ and the field strength $\SF$ of $\SU(2)_R$ multiplied by $1/2\pi $ and $i/2\pi$ respectively.
We also define
\beq
\tilde{\hat{\SR}} = \left(\frac{1}{2}\hat{\SR}_{kl} \epsilon_{kl ij} \right)_{ij}.
\eeq
%Furthermore, define 
%\beq
%\hat{W} &=\frac{1}{2 \pi} \left( \frac{1}{2}W_{ij \mu \nu} dx^\mu \wedge dx^\nu  \right)_{ij} 
%\eeq 
%to be the analog of $\hat{R}$ by replacing the Riemann curvature by the Weyl tensor.
Topological terms are
\beq
e & :=  -  \frac{1}{4} \tr ( \hat{\SR} \tilde{\hat{\SR}} ) = \frac{1}{2^5} \hat{\SR}_{ij \mu\nu} {\hat \SR}_{kl \rho \sigma} \epsilon_{ijkl}\epsilon_{\mu\nu\rho\sigma}, \\
p_1 & :=  -\frac{1}{2} \tr ( \hat{\SR} \hat{\SR} ) =  \frac{1}{2^3} \hat{\SR}_{ij \mu\nu} \hat{\SR}_{ij \rho \sigma} \epsilon_{\mu\nu\rho\sigma} ,  \\
c_2 & := - \frac{1}{2} \tr \hat{\SF}^2,  
\eeq
which are used in the main text.
Non-topological terms are
\beq
[\SR_{\mu\nu\rho\sigma}^2]  & := -  \frac{1}{2} \tr ( \hat{\SR} \wedge \star \hat{\SR} ) = \frac{1}{2^2} \hat{\SR}_{ij \mu\nu} \hat{\SR}_{ij \mu \nu},   \\
[\SR^2]  & :=  \hat{\SR}^2 , \\
[\SF^2] & :=  - \frac{1}{2} \tr ( \hat{\SF} \wedge \star \hat{\SF} ), \\
[\SD^2] & := \hat{\SD}^2.
\eeq
where $\star$ is the Hodge star, and $\hat{\SR}=\SR/2\pi$ and $\hat{\SD}=\SD/2\pi$.

%For later purposes, let us also present some relations. First, it is easy to check that $p_1$ can be represented just by the Weyl tensor as
%\beq
%p_1 &=  -\frac{1}{2} \tr ( \hat{W} \hat{W} ) =  \frac{1}{2^4} \hat{W}_{ij \mu\nu} \hat{W}_{ij \rho \sigma} \epsilon_{\mu\nu\rho\sigma}   
%\eeq
Let $\hat{\SF}_\ell$ and $\hat{\SF}_r$ be the curvature multiplied by $\frac{i}{2\pi}$
for the spin connections of $\SU(2)_\ell$ and $\SU(2)_r$ Lorentz group, respectively. Then we have
\beq
e &=   \frac{1}{2} \tr ( \hat{\SF}_\ell^2 ) -  \frac{1}{2} \tr ( \hat{\SF}_r^2 ), \\
p_1 &=   \tr ( \hat{\SF}_\ell^2 ) +  \tr ( \hat{\SF}_r^2 )   , \\
[\SR_{\mu\nu\rho\sigma}^2]  &= \tr ( \hat{\SF}_\ell \wedge \star \hat{\SF}_\ell ) +  \tr ( \hat{\SF}_r \wedge \star \hat{\SF}_r ).
\eeq

Now let us start the analysis of allowed terms.
First, notice that $[\SR_{\mu\nu\rho\sigma}^2] ,[\SR^2], [\SF^2]$ and $[\SD^2]$ are not topological invariant.
However, after the topological twist $\SF \to \SF_r$ and $\SD \to \alpha \SR$, the effective action must be described by topological invariants.
The only combinations of them to give topological invariants are
\beq
[\SR_{\mu\nu\rho\sigma}^2] + 4 [\SF^2], \qquad\text{or}\qquad
[\SD^2] - \alpha^2[\SR^2].
\eeq
The first is a topological invariant after the topological twist since
\begin{equation}
 \tr ( \hat{\SF}_\ell \wedge \star \hat{\SF}_\ell ) -  \tr ( \hat{\SF}_r \wedge \star \hat{\SF}_r ) 
=  -  \frac{1}{2} \tr ( \hat{\SR} \wedge \star \tilde{\hat{\SR}} ) 
= \frac{1}{2^3} \hat{\SR}_{ij \mu\nu} \hat{\SR}_{kl \mu \nu} \epsilon_{ijkl} 
=p_1.
\end{equation}
The second  just vanishes after the topological twist. So the effective action must be of the form
\beq
- \CL  &\ni ( f + \bar{f})e + ( g - \bar{g}) p_1 + ( h - \bar{h}) c_2 \nonumber \\
& + ( j +\bar{j}) ([\SR_{\mu\nu\rho\sigma}^2) +4 [\SF^2]] + (k +\bar{k})( [\SR^2] - \alpha^{-2} [\SD^2]  ).
\eeq
After the topological twist $\SF \to \SF_r$, we have
\beq
c_2  \to \frac{1}{2}e - \frac{1}{4}p_1
\eeq
and hence the effective action becomes
\beq
\left( (f+\bar{f}) +\frac{1}{2}( h - \bar{h}) \right)e +\left(( g - \bar{g}) - \frac{1}{4}( h - \bar{h}) + ( j +\bar{j})  \right)p_1.
\eeq
This must be holomorphic~\cite{Witten:1995gf}, and hence we get
\beq
\bar{f} - \frac{1}{2} \bar{h}=0,~~~~~ - \bar{g} + \frac{1}{4}  \bar{h}  +\bar{j}=0.
\eeq
Therefore, the possible form of the effective action is given as
\beq
- \CL  &\ni ( f + \bar{f})e + ( g - \bar{g}) p_1 + 2( f - \bar{f}) c_2 \nonumber \\
& + ( -\frac{f+\bar{f}}{2} +g +\bar{g}) ([\SR_{\mu\nu\rho\sigma}^2] +4 [\SF^2]) + (k +\bar{k}) ( [\SR^2] - \alpha^{-2}[\SD^2] ).
\eeq
%\beq
%- \CL  &\ni  f \left( E + 2c_2 - \frac{1}{2} [[\SR_{\mu\nu\rho\sigma}^2] +4 [\SF^2]]  \right) \nonumber \\
%&+ g \left( p_1 +  [[\SR_{\mu\nu\rho\sigma}^2] +4 [\SF^2]]  \right) \nonumber \\
%&+  \bar{f} \left( E - 2c_2 - \frac{1}{2} [[\SR_{\mu\nu\rho\sigma}^2] +4 [\SF^2]]  \right) \nonumber \\
%&+ \bar{g} \left( - p_1 +  [[\SR_{\mu\nu\rho\sigma}^2] +4 [\SF^2]]  \right)
% \label{eq:allowed}
%\eeq

It may be more familiar in the context of conformal field theory to rewrite $[\SR_{\mu\nu\rho\sigma}^2]$ in terms 
of Weyl squared,
\beq
[\SC_{\mu\nu\rho\sigma}^2] & := -  \frac{1}{2} \tr ( \hat{\SC} \wedge \star \hat{\SC} ) = \frac{1}{2^2} \hat{\SC}_{ij \mu\nu} \hat{\SC}_{ij \mu \nu}   
\eeq
where $ \hat{\SC}_{ij \mu \nu} =\SC_{ij \mu \nu}/2\pi $ is the Weyl tensor divided by $2\pi$,
\beq
 \hat{\SC}_{ijkl} =  \hat{\SR}_{ij kl} - \frac{1}{2}( \delta_{ik}\hat{\SR}_{jl} - \delta_{il}\hat{\SR}_{jk} -\delta_{jk}\hat{\SR}_{il}+\delta_{jl}\hat{\SR}_{ik} ) 
 +\frac{1}{6}(\delta_{ik}\delta_{jl}-\delta_{il}\delta_{jk})\hat{\SR}.
\eeq
A little algebra gives $[\SR_{\mu\nu\rho\sigma}^2]  = 2[\SC_{\mu\nu\rho\sigma}^2]-2e + \frac{1}{12}[\SR^2]$.
%\beq
% \hat{W}_{ijkl} =  \hat{R}_{ij kl} - \frac{1}{d-2}( \delta_{ik}\hat{R}_{jl} - \delta_{il}\hat{R}_{jk} -\delta_{jk}\hat{R}_{il}+\delta_{jl}\hat{R}_{ik} ) 
% +\frac{1}{(d-1)(d-2)}(\delta_{ik}\delta_{jl}-\delta_{il}\delta_{jk})\hat{R},
%\eeq
%where $d=4$. 
\if0
We have
\beq
(\hat{W}_{\mu\nu\rho\sigma}^2)  & = \frac{1}{2^2} \hat{W}_{ij kl} \hat{W}_{ijkl}   \nonumber \\
& =  \frac{1}{2^2} \left( \hat{R}_{ij kl} \hat{R}_{ijkl}  - 2\hat{R}_{ij}\hat{R}_{ij} + \frac{1}{3}\hat{R}^2 \right)
\eeq
On the other hand,
\beq
E &=  \frac{1}{2^5} \hat{R}_{ij \mu\nu} {\hat R}_{kl \rho \sigma} \epsilon_{ijkl}\epsilon_{\mu\nu\rho\sigma} \nonumber \\
&= \frac{1}{2^3}(  \hat{R}_{ij kl} \hat{R}_{ijkl}  - 4\hat{R}_{ij}\hat{R}_{ij}  +  \hat{R}^4 )
\eeq
From them, we get
\beq
[\SR_{\mu\nu\rho\sigma}^2]  & = \frac{1}{2^2} \hat{R}_{ij kl} \hat{R}_{ijkl}   \nonumber \\
& = 2[\SC_{\mu\nu\rho\sigma}^2]-2E + \frac{1}{12}[\SR^2]
\eeq
\fi
Substituting this to the action
\if0
, we get
\beq
- \CL  \ni
&  2( (f + \bar{f}) - (g + \bar{g} ) )E  + ( g - \bar{g}) p_1 + 2( f - \bar{f}) c_2 \nonumber \\
&+ ( -\frac{f+\bar{f}}{2} +g +\bar{g}) [ 2[\SC_{\mu\nu\rho\sigma}^2]+ \frac{a^{-2}}{12}[\SD^2]+4 [\SF^2]]  \nonumber \\
&+ (k +\bar{k}-\frac{f+\bar{f}}{24} +\frac{g +\bar{g}}{12})[ [\SR^2] - a^{-2}[\SD^2] ] 
\eeq
To simplify it, set
\fi
and redefining the coefficient functions as
\beq
F &:= 2f-2g, &
G & := -f+2g, &
H & := k -\frac{f}{24} +\frac{g }{12}
\eeq
%from which we get $f=F+G$ and $g=\frac{1}{2}F+G$. Then the effective action becomes
we get
\beq
- \CL  \ni & F \left(e +\frac{1}{2}p_1+ 2c_2 \right) \nonumber \\
  + & G \left( [\SC_{\mu\nu\rho\sigma}^2]+ \frac{\alpha^{-2}}{24}[\SD^2]+2 [\SF^2] + p_1+2c_2  \right) \nonumber \\
  +& H \left([\SR^2] - \alpha^{-2}[\SD^2]   \right) \nonumber \\
  + & \text{anti-holomorphic terms}. \label{eq:allowed}
\eeq
The $F$ and $G$ can be taken to be arbitrary, or otherwise the analysis of \cite{Witten:1995gf} would fail.
At the level of the analysis presented here, it is not clear whether $H$ can be taken arbitrary or not.

\paragraph{Check: the conformal anomaly}
Let us perform a check of the above result.
In the current normalization, the conformal anomaly is given as
\beq
T^\mu_\mu = c [\SC_{\mu\nu\rho\sigma}^2]- 2a e + \cdots \label{eq:confanom}
\eeq
where $c$ and $a$ are the conformal central charges normalized in such a way that
$n_h$ hypermultiplets and $n_v$ vector multiplets contribute as
$c=\frac{1}{12}n_h + \frac{1}{6}n_v $ and $a=\frac{1}{24}n_h + \frac{5}{24}n_v$. 
The ellipsis represents contributions from other background fields $\SF$ and $\SD$.

The anomaly of dilatation and $\U(1)_R$ are known to combine into a holomorphic term.
This can be seen as follows. The combination $J^\mu:=J^\mu_D +  \frac{i}{2} J_R^\mu$ of the dilatation current $J^\mu_D :=T^{\mu\nu} x_\mu$ and the $\U(1)_R$ current $J_R^\mu$
acts trivially on anti-holomorphic Coulomb branch operators because of the relation $R+2D=0$ between the scaling dimension $D$ and the $\U(1)_R$ charge $R$. 
The anomaly matching of these symmetries on a generic point of the Coulomb branch is done by an effective action of the form \eqref{eq:allowed},
and hence the anomaly of $\partial_\mu J^\mu=T^\mu_\mu + \frac{i}{2} \partial_\mu J_R^\mu$ is saturated by the holomorphic terms.
We get
\begin{multline}
T^\mu_\mu + \frac{i}{2} \partial_\mu J_R^\mu = 
  c  \left( [\SC_{\mu\nu\rho\sigma}^2]+ \frac{\alpha^{-2}}{24}[\SD^2]+2 [\SF^2] + p_1+2c_2  \right)-
2a \left(e +\frac{1}{2}p_1+ 2c_2 \right) ,
\end{multline}
where the coefficients are chosen to match the conformal anomaly \eqref{eq:confanom}.
There is a priori reason that the coefficients of $ [\SF^2] $ and $[\SD^2]$ are proportional to $c$.
They both come from the scaling anomaly of the two point function of the energy-momentum multiplet.

For $\U(1)_R$, we get
\beq
 i\partial_\mu J_R^\mu &= 2(c-a)p_1 + 2(2c -4a) c_2 \nonumber \\
 &= (c-a)  \frac{1}{2^2} \hat{\SR}_{ij \mu\nu} \hat{\SR}_{ij \rho \sigma} \epsilon_{\mu\nu\rho\sigma} 
 - (c -2a) \frac{1}{2} \tr \hat{\SF}_{\mu\nu} \hat{\SF}_{\rho\sigma}\epsilon_{\mu\nu\rho\sigma}.
\eeq
The imaginary factor $i$ appears because $\epsilon_{\mu\nu\rho\sigma} $ is the totally anti-symmetric tensor in Euclidean space
and hence it gets a factor $i$ when rotated back to Minkowski space.
This equation agrees with the standard anomaly equation of $\U(1)_R$; see e.g. Sec.~2 of \cite{Shapere:2008zf}.
Putting $(c-a)=(n_h-n_v)/24$ and $2c-4a=-n_v/2$ gives $ i\partial_\mu J_R^\mu = n_h  (2p_1/24) - n_v (2p_1/24+c_2) $ which is exactly as expected from the Atiyah-Singer index theorem.

\bibliographystyle{ytphys}
\bibliography{ref}

\end{document}